\documentclass[12pt]{article}
\usepackage[dvips]{feynmp} 
\begin{fmffile}{muon_figury}

\newcommand{\be}{\begin{eqnarray}}
\newcommand{\ee}{\end{eqnarray}}
\def\dd{{\rm d}}
\def\vep{\epsilon}

\begin{document}\setlength{\unitlength}{1mm}

\thispagestyle{empty}

\vspace*{20mm}

\begin{flushright}
{\bf BNL-HET-98/38}\\
{\bf hep-ph/9812394}
\end{flushright}
\vspace{0.6cm}
\boldmath
\begin{center}
\Large\bf\boldmath The muon anomalous magnetic moment in QED:
three-loop electron and tau contributions
\end{center}
\unboldmath
\vspace{0.8cm}

\begin{center}

{\large Andrzej Czarnecki}\\
{\sl Physics Department, Brookhaven National Laboratory,\\
Upton, NY 11973, USA}

\vspace*{5mm}

{\large Maciej Skrzypek}\\
{\sl Institute of Nuclear Physics,
  ul. Kawiory 26a, 30-055 Cracow, Poland}\\
 and\\
{\sl CERN, Theory Division, CH-1211 Geneva 23, Switzerland}
\end{center}

\vfill

\begin{abstract}
We present an analytic calculation of electron and tau ${\cal
O}(\alpha^3)$ loop effects on the muon anomalous magnetic moment.
Computation of such three-loop diagrams with three mass scales is
possible using asymptotic and eikonal expansions.  An evaluation of a
new type of eikonal integrals is presented in some detail.
\end{abstract}

\vfill

\newpage
\section{Introduction}
The new measurement of the muon anomalous magnetic moment,
$a_\mu=(g_\mu-2)/2$, in the experiment E821 in Brookhaven has motivated
many recent theoretical studies.  In the Standard Model, $a_\mu$
receives contributions from electromagnetic and weak interactions, as
well as from loop effects involving hardrons.  All three types of
effects have been studied recently.  QED contributions are known to
the four-loop level, and even some five-loop diagrams have been
evaluated.  The next largest contribution is due to hadronic loops and
is the most difficult one to evaluate.  There has been recently
significant progress in both evaluation of the light-by-light diagrams
\cite{Bijnens:1996xf} and hadronic vacuum polarization effects
\cite{Jeg95}.  Electroweak two-loop effects are also known
\cite{CKM96}.

\begin{figure}[h]
\begin{center}
  \begin{fmfgraph*}(40,40)
    \fmfbottomn{i}{1}\fmftopn{o}{2}
    \fmflabel{$\gamma$}{i1}
    \fmflabel{$\mu$}{o1}
    \fmflabel{$\mu$}{o2}
    \fmf{fermion}{o1,v1}
    \fmf{plain}{v1,tt1}
    \fmf{fermion}{tt1,tt3}
    \fmf{plain}{tt3,v2,tt4}
    \fmf{fermion}{tt4,tt2}
    \fmf{plain}{tt2,v3}
    \fmf{fermion}{v3,o2}
    \fmf{photon}{i1,v2}
    \fmffreeze
    \fmf{photon}{v7,v1}
    \fmf{photon}{v6,v5}
    \fmf{photon}{v4,v3}
    \fmf{fermion,left}{v4,v5}
    \fmf{fermion,left}{v6,v7}
    \fmffreeze
    \fmf{fermion,left,label=$e$}{v5,v4}
    \fmf{fermion,left,label=$\tau$}{v7,v6}
  \end{fmfgraph*}
\end{center}
\caption{\sf Three-loop contribution to muon anomalous magnetic moment
with the electron and tau loops inserted in the photon propagator.}
\label{fig:threeloop}
\end{figure}
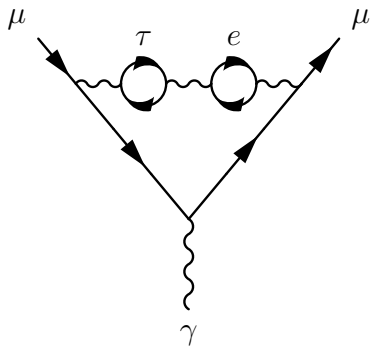

The present paper is devoted to the only 3-loop QED contribution to
$a_\mu$ that has not been evaluated analytically so far:  from 
a diagram
with electron and $\tau$ lepton loop insertions in the photon
propagator, shown in Fig.~\ref{fig:threeloop}.
Because of the three mass scales present in this diagram,
$m_{e,\mu,\tau}$, it cannot at present be evaluated in a
closed form.  However, we present an approach based on asymptotic and
eikonal expansions, which takes advantage of the wide separations
between those scales and permits an evaluation with an arbitrary
accuracy.

We begin with a brief summary of the present knowledge of QED
contributions to the electron and muon anomalous magnetic moments,
$a_{e,\mu}$ (for more details see e.g.~\cite{Czarnecki:1998nd}).

{\bf (a) Electron.}
To match the present experimental precision, one needs 
four terms in the expansion of $a_e$ in the fine 
structure constant $\alpha$,
\be
a_e = \sum_{n=1}^4 A_n \left( {\alpha\over \pi}\right)^n
 + \ldots
\ee
where ellipses indicate
contributions of loops containing the heavy leptons $\mu$ and
$\tau$; $A_{1,2}$ have been known since the early years of QED
\cite{Schwinger48}.  An analytical
evaluation of $A_3$ required the efforts of many groups and took almost 40
years; it has been completed only recently \cite{Laporta:1996mq}. $A_4$ is
known only numerically.

The perturbative series for $a_e$ is very well behaved,
and, together with the most recent experimental values for electron and
positron \cite{Dehmelt87}, 
allows one to deduce a very precise value of the fine structure
constant \cite{Dubna,Czarnecki:1998nd},
\be
\alpha^{-1} = 137.03599959(38)(13).
\label{eq:alpha}
\ee

{\bf (b) Muon.}  The value of $\alpha$ found from electron $g-2$,
eq.~(\ref{eq:alpha}),  can be applied to compute the QED contribution
to the muon anomalous magnetic moment $a_\mu$. 
Because of the presence of electron loops, higher-order QED contributions
to $a_\mu$ are enhanced with respect to $a_e$.  At present
five terms of the expansion in $\alpha$ are needed:
\be
a_\mu^{\rm QED} = \sum_{n=1}^5 C_n\left( {\alpha\over \pi}\right)^n
\ee
with
\be
C_1 &=& A_1 = 0.5,
\nonumber \\
C_2 &=& A_2 + a_1(m_e/m_\mu) + a_2(m_\mu/m_\tau) =
0.765\;857\;388(44),
\nonumber \\
C_3 &=& A_3+C_3^{\gamma\gamma}(e)+C_3^{\gamma\gamma}(\tau)
 +C_3^{\rm vac.\ pol.}(e) +C_3^{\rm vac.\ pol.}(\tau)
 +C_3^{\rm vac.\ pol.}(e,\tau) 
\nonumber \\
&=& 24.050\;509(2),
\nonumber \\
C_4 &=& A_4+ 127.55(41) = 126.04(41),
\nonumber \\
C_5&=&  930(170),
\ee
where $a_{1,2}$ describe contributions of 
two-loop diagrams with electron and tau
loops, respectively \cite{Li:1993xf,Elend66}.  We will discuss them in
detail later on. 

In $C_3$ we have contributions from light-by-light scattering diagrams
with $e$ and $\tau$ loops \cite{Laporta:1993pa}, 
and vacuum polarization diagrams with
either $e$, or $\tau$ \cite{Laporta:1993ju}, 
or both types of loops.  
An analytical evaluation of this latter contribution 
of mixed $e-\tau$ diagram, 
$C_3^{\rm vac.\ pol.}(e,\tau)$, is the main purpose of this paper and 
will be presented below. Numerically it can be evaluated using 
the kernel from \cite{Krause:1997rf}: 
\be C_3^{\rm vac.\ pol.}(e,\tau) = 0.0005276(2) .
\ee
For $C_4$ one uses the difference
between the muon and electron coefficients found in
\cite{Kinoshita:1993pq}. For $C_5$ only a numerical estimate of the
presumably dominant contributions is known \cite{Karshenboim:1993rt}.

The present estimate of the total QED contribution to $a_\mu$ is
\cite{Czarnecki:1998nd}
\be
a_\mu^{\rm QED} = 116584705.6(2.9)\times 10^{-11}. 
\ee

\section{\boldmath 
Two-loop diagrams with electron and tau loops}

\begin{figure}[h]
\begin{center}
  \begin{fmfgraph*}(40,40)
    \fmfbottomn{i}{1}\fmftopn{o}{2}
    \fmflabel{$\gamma$}{i1}
    \fmflabel{$\mu$}{o1}
    \fmflabel{$\mu$}{o2}
    \fmf{fermion}{o1,v1}
    \fmf{plain}{v1,tt1}
    \fmf{fermion}{tt1,tt3}
    \fmf{plain}{tt3,v2,tt4}
    \fmf{fermion}{tt4,tt2}
    \fmf{plain}{tt2,v3}
    \fmf{fermion}{v3,o2}
    \fmf{photon}{i1,v2}
    \fmffreeze
    \fmf{photon,label=$\gamma$}{v5,v1}
    \fmf{photon,label=$\gamma$}{v3,v4}
    \fmf{fermion,left}{v4,v5}
    \fmffreeze
    \fmf{fermion,left,label=$e,,\mu,,\tau$}{v5,v4}
  \end{fmfgraph*}
\end{center}
\caption{\sf Two-loop contributions of lepton loops in the photon
propagator.}
\label{fig:one}
\end{figure}
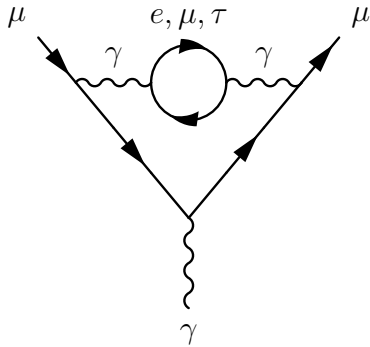

Before calculating the contribution of the 
three-loop diagram of Fig.~\ref{fig:threeloop},
we would like to discuss a method of evaluating the
two-loop diagrams shown if Fig.~\ref{fig:one} 
(we discuss only the Pauli formfactor, relevant for $a_\mu$).
It will serve us as an example to illustrate the main points of our
calculational techniques.
%
Of course, full analytical
results for these two-loop contributions are known  
\cite{Li:1993xf,Elend66}.  It is
instructive to note how they were obtained.  First, a closed
analytical formula was found, with full functional dependence on the
ratio of masses $m_e/m_\mu$ or $m_\mu/m_\tau$ \cite{Elend66}.  That
formula, containing dilogarithms, was found awkward to use, because
of cancellations and difficulties in estimating the accuracy
in the numerical evaluation.  In \cite{Li:1993xf}  expansions of the
exact result in powers of small mass ratios were given.  Such
expansions avoid evaluation of special functions and their accuracy
can be precisely assessed.  

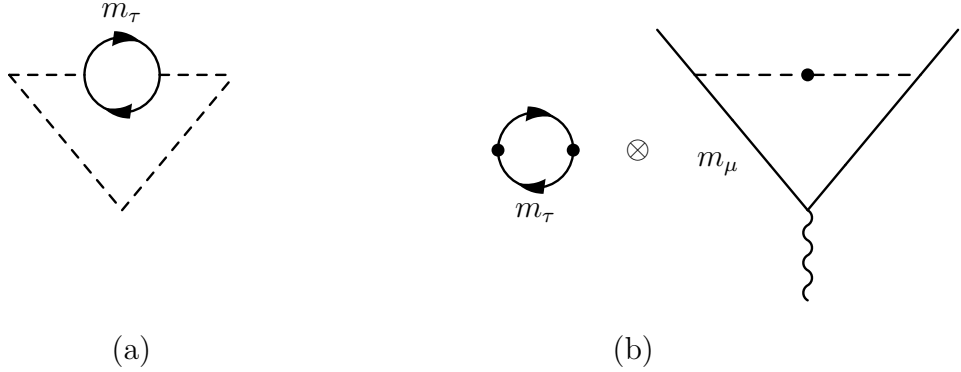
\begin{figure}[h]
\begin{displaymath}
  \parbox{70mm}{ 
  \begin{fmfgraph*}(40,40)
    \fmfbottomn{i}{1}\fmftopn{o}{2}
    \fmf{phantom}{o1,t1}
    \fmf{dashes}{t1,t3,t5,j1,t7,t4,t2}
    \fmf{phantom}{t2,o2}
    \fmf{phantom}{i1,j1}
    \fmffreeze
    \fmf{dashes}{t1,v5}
    \fmf{dashes}{v4,t2}
    \fmf{fermion,left}{v4,v5}
    \fmffreeze
    \fmf{fermion,left,label=$m_\tau$}{v5,v4}
  \end{fmfgraph*}}
%
  \parbox{16mm}{ 
  \begin{fmfgraph*}(10,10)
    \fmfleftn{i}{1}\fmfrightn{o}{1}
    \fmf{fermion,left,tension=2}{i1,o1}
    \fmf{fermion,left,label=$m_\tau$}{o1,i1}
    \fmfdot{i1,o1}
  \end{fmfgraph*}}
\otimes
  \parbox{40mm}{ 
  \begin{fmfgraph*}(40,40)
    \fmfbottomn{i}{1}\fmftopn{o}{2}
    \fmf{plain}{o1,t1,t3}
    \fmf{plain,label=$m_\mu$}{t5,t3}
    \fmf{plain}{t5,j1,t7,t4,t2,o2}
    \fmf{photon}{i1,j1}
    \fmffreeze
    \fmf{dashes}{t1,v1,t2}
    \fmfdot{v1}
  \end{fmfgraph*}}
\end{displaymath}
{\hskip 2.8truecm (a) \hskip 6truecm (b)}
\caption{\sf Graphic representation of characteristic momentum scales in
the two-loop diagram with a $\tau$ loop.  Solid and dashed lines
denote, respectively, massive and massless 
propagators.}
\label{fig:tau}
\end{figure}

Here we demonstrate how such expansions can be obtained {\em without}
knowledge of the exact result.  We first consider the case of the
$\tau$ loop insertion.  In this case a well known method of heavy mass
expansion is applicable (for a review see
\cite{Chetyrkin91}).  Let us denote the
loop momenta by $p_\tau$ and $p_\mu$, for the momentum
flowing inside the $\tau$ loop and for that in the virtual photon 
respectively.
There are two regions of integration, with characteristic scales of
momenta $p_\tau \sim p_\mu \sim m_\tau$ and $p_\tau \sim m_\tau$,
$p_\mu \sim m_\mu$.  In the case of the first region, which we can call
the hard contribution, we can safely regard the muon mass and external
momentum as small with respect to the integration momenta, and expand the
integrand in these small parameters.  The resulting integral
corresponds to a vacuum diagram shown in Fig.~\ref{fig:tau}(a).  
In the second region we cannot neglect the external muon momentum, but
now, since $p_\mu \ll p_\tau$, we can expand the $\tau$ loop
propagators in Taylor series in $p_\mu$, so that the integral
factorizes into a product of two one-loop diagrams, shown in
Fig.~\ref{fig:tau}(b).  This nice factorization of the relevant
integration regions is only possible in regularization schemes, which
do not introduce additional mass scales, such as dimensional
regularization.  

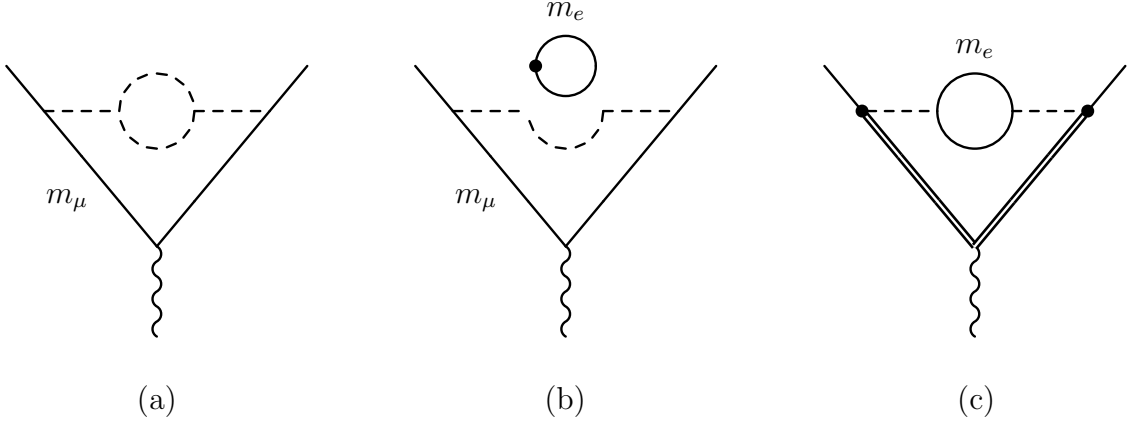
\begin{figure}[h]
\vskip 0.5truecm
\begin{center}
  \begin{fmfgraph*}(40,40)
    \fmfbottomn{i}{1}\fmftopn{o}{2}
    \fmf{plain}{o1,t1,t3}
    \fmf{plain,label=$m_\mu$}{t5,t3}
    \fmf{plain}{t5,j1,t7,t4,t2,o2}
    \fmf{photon}{i1,j1}
    \fmffreeze
    \fmf{dashes}{t1,v5}
    \fmf{dashes}{v4,t2}
    \fmf{dashes,left}{v4,v5}
    \fmffreeze
    \fmf{dashes,left}{v5,v4}
  \end{fmfgraph*}
\hskip 1.3truecm
  \begin{fmfgraph*}(40,40)
    \fmfbottomn{i}{1}\fmftopn{o}{2}
    \fmf{plain}{o1,t1,t3}
    \fmf{plain,label=$m_\mu$}{t5,t3}
    \fmf{plain}{t5,j1,t6,t4,t2,o2}
    \fmf{photon}{i1,j1}
    \fmffreeze
    \fmf{dashes}{t1,v5}
    \fmf{dashes}{v4,t2}
    \fmf{dashes,left}{v4,v5}
    \fmf{phantom}{o1,vi5,vi4,o2}
    \fmf{plain,left,label=$m_e$}{vi5,vi4}
    \fmffreeze
    \fmf{plain,left}{vi4,vi5}
    \fmfdot{vi5}
  \end{fmfgraph*}
\hskip 1.3truecm
  \begin{fmfgraph*}(40,40)
    \fmfbottomn{i}{1}\fmftopn{o}{2}
    \fmf{plain}{o1,t1}
    \fmf{dbl_plain}{t1,t3,t5,j1,t6,t4,t2}
    \fmf{plain}{t2,o2}
    \fmf{photon}{i1,j1}
    \fmffreeze
    \fmf{dashes}{t1,v5}
    \fmf{dashes}{v4,t2}
    \fmf{plain,left}{v4,v5}
    \fmffreeze
    \fmf{plain,left,label=$m_e$}{v5,v4}
    \fmfdot{t1,t2}
  \end{fmfgraph*}
\end{center}
{\hskip 1.8truecm (a) \hskip 4.75truecm (b) \hskip 4.8truecm (c)}
\caption{\sf Integration regions contributing to the two-loop diagram with
an electron loop.  (a): Taylor expansion in $m_e$.  (b): region of
soft electron loop momentum.  (c): region of both momenta soft (double
line denotes muon propagator expanded in the square of its virtual
momentum).} 
\label{fig:e}
\end{figure}

After evaluation of these two sub-diagrams and renormalization of the
electric charge of the muon, we find a finite result.  Several of the terms
we calculated agree with formula (12) in
\cite{Li:1993xf}:
\begin{eqnarray}
a_2&&\!\!\!\!\!\!\!\!\!(l = m_\mu/m_\tau)=
\frac{l^2}{45}+
\frac{l^4\ln l}{70}
+\frac{9}{19600}l^4-
\frac{ 131 }{ 99225 } l^6
+ \frac{4l^6}{315}\ln l
\nonumber\\
&&
-\sum^\infty_{n=3}
\frac{(8n^3+28n^2-45) l^{2n+2}}
     {[(n+3)(2n+3)(2n+5)]^2}
+2 \ln l
\sum^\infty_{n=3}
\frac{nl^{2n+2}}{(n+3)(2n+3)(2n+5)}.
\nonumber
\end{eqnarray}
The other two-loop diagram that we have to consider is the 
electron loop insertion in the photon propagator (Fig.~\ref{fig:one}).  
Here the situation is somewhat more involved, since there are now three
integration regions.  Introducing the obvious notation for the
integration momenta, $p_{e,\mu}$, we have: ($p_e\sim p_\mu\sim m_\mu$),
($p_e\sim m_e$ and $p_\mu\sim m_\mu$), and ($p_e\sim p_\mu\sim m_e$).
The first two regions correspond to known cases of the ``large
momentum expansion''
\cite{Chetyrkin91}.  They are depicted
in Fig.~\ref{fig:e}(a,b).  
Figure~\ref{fig:e}(a) denotes a simple Taylor
expansion of the electron propagators in $m_e$, justified if both
integration momenta are large.  In Fig.~\ref{fig:e}(b) we have
illustrated one of
the two cases, where one of the electron lines cannot be expanded in
$m_e$, but where the integration factorizes and we have a product of
one-loop diagrams.  More exotic is the third case, Fig.~\ref{fig:e}(c), 
where the only scale
of integration is $m_e$, since all dependence on the muon mass and
external momentum factorizes.  Sub-diagrams of this type have been
encountered in a different context in eikonal expansions
\cite{CzarSmir96,Smirnov96}.  Integrals that arise in the present
case are somewhat different and we give here some details of their
evaluation.  

Since now the integration momentum $p_\mu$ is much smaller than the
external muon momentum $p$, we can expand muon propagators in $p_\mu^2$.
As a result we have to compute integrals of the form
\be
J(a_1,a_2) ={1\over \pi^D} \int {
\dd^D p_\mu \dd^D p_e
\over
(p_\mu^2)^{a_1}
(2p_\mu\cdot p)^{a_2}
(p_e^2+m_e^2)
[(p_\mu+p_e)^2+m_e^2]
}.
\label{eq:J}
\ee
First, we use Feynman parameters
to combine the last two propagators and integrate over $p_e$.  We get
\be
J(a_1,a_2) ={1\over \pi^{D/2}}\Gamma(\vep) x^{-\vep}(1-x)^{-\vep}
\int {
\dd^D p_\mu 
\over
(p_\mu^2)^{a_1}
(2p_\mu\cdot p)^{a_2}
(p_\mu^2+m_x^2)^{\vep}
},
\ee
with $m_x^2 \equiv m_e^2/x(1-x)$.  Next, using again Feynman
parameters, we combine the first and the last term in the denominator.
Finally, using
\be
{1\over a^\alpha b^\beta} 
={1\over B(\alpha,\beta)}
\int_0^\infty {\rm d}\lambda \,
{\lambda^{\beta-1} \over 
   \left[ a+b\lambda \right]^{\alpha+\beta}},
\label{eq:geor}
\ee
and integrating over $p_\mu$, we get
\be
\lefteqn{J(a_1,a_2)= {\Gamma\left({a_2\over 2}\right)
\Gamma\left(a_1+{a_2\over 2}-2+2\vep \right)
\Gamma\left(2-a_1-{a_2\over 2}-\vep\right)
\Gamma^2\left(-1+a_1+{a_2\over 2}+\vep\right)
\over 
2\Gamma(a_2) 
\Gamma\left(2-{a_2\over 2}-\vep\right)
\Gamma(-2+2a_1+a_2+2\vep)
}
}
\nonumber \\
&& \qquad \qquad \times
{m_e^{4-2a_1-a_2-4\vep}\over m_\mu^{a_2}}
\nonumber \\
&=& 
{-\pi^2 \Gamma\left(a_1+{a_2\over 2}-2+2\vep \right)
m_e^{4-2a_1-a_2-4\vep}
\over 
2^{2a_1+2a_2-3+2\vep}
\Gamma\left( {a_2+1\over 2}\right)
\Gamma\left(a_1+{a_2-1\over 2}+\vep \right)
\Gamma\left(2-{a_2\over 2}-\vep \right)
\sin\pi\left( a_1+{a_2\over 2}+\vep\right)
 m_\mu^{a_2}}.
\nonumber 
\ee
We should mention that the integrand in eq.~(\ref{eq:J}) could also
contain products $p\cdot p_e$.  It is possible to replace them by
combinations of products of $p_\mu\cdot p_e$ and $p_\mu\cdot p$ using
traceless products (see e.g.~\cite{CzarSmir96}).  However, in the
present case there are at most two powers of
$p_e$ in the numerator and we can use the following simple formulas:
\be
(p\cdot p_e)^2 
\to
{p^2 p_e^2 \over D}+\left[
(p_\mu\cdot p)^2-{p^2 p_\mu^2 \over D}\right]
{
D(p_e\cdot p_\mu)^2-p_e^2 p_\mu^2
\over 
(D-1)(p_\mu^2)^2
},\;\;\;
p\cdot p_e 
\to {(p\cdot p_\mu)( p_\mu\cdot p_e)\over p_\mu^2}.
\nonumber
\ee
Adding the contributions of the three integration regions, we find,
after renormalization, that the terms we obtained agree 
with formula (11) of \cite{Li:1993xf}:
\begin{eqnarray}
a_1(k=m_e/m_\mu)&=&
-\frac{25}{36}
+\frac{\pi^2}{4}k
-\frac{1}{3}\ln k
+(3+4\ln k)k^2
-\frac{5}{4}\pi^2k^3
\nonumber\\
&&
+
\left[\frac{\pi^2}{3}
+\frac{44}{9}
-\frac{14}{3} \ln k
+ 2 \ln^2k
\right]k^4
+\frac{8}{15}k^6\ln k
-\frac{109}{225}k^6
\nonumber\\
&&
+\sum^\infty_{n=2}
\left[
\frac{2(n+3)}{n(2n+1)(2n+3)}\ln k
-\frac{8n^3+44n^2+48n+9}{n^2(2n+1)^2(2n+3)^2}
\right]
k^{2n+4}.
\nonumber
\end{eqnarray}

\section{\boldmath 
  Three-loop diagram with $e$ and $\tau$ loop insertions}

We now proceed to the actual focus of our work, the contribution of 
the three-loop diagram
with $e$ and $\tau$ loop insertions, shown in Fig.~\ref{fig:threeloop}.
So far its contribution to $a_\mu$ has been evaluated only numerically
\cite{Laporta:1993ju}.%

Using the techniques described above, we can easily obtain an expansion
of the Pauli part of this diagram with arbitrary accuracy.
  There are now three integration
momenta and we have to consider five integration regions, combinations of
the conditions described in the context of two-loop diagrams.  
Using the notation $p_{\tau,\mu,e}$ for the integration momenta in the
three loops, the regions we have to consider are:
\begin{itemize}
\item $p_\tau \sim p_\mu \sim p_e \sim m_\tau$,
\item $p_\tau \sim m_\tau$; $p_\mu \sim p_e \sim m_\mu$,
\item $p_\tau \sim m_\tau$; $p_\mu \sim m_\mu$; $p_e \sim m_e$,
\item $p_\tau \sim m_\tau$; $p_\mu \sim p_e \sim m_e$,
\item $p_\tau \sim p_\mu \sim m_\tau$; $p_e \sim m_e$.
\end{itemize}
Calculations in each of these regions are analogous to the cases
described in the previous section. 
For
present purposes it is more than sufficient to retain the first 
three terms in the $m_\mu^2/m_\tau^2$ expansion and two terms in
$m_e^2/m_\tau^2$.  After renormalization we find 
\be
\lefteqn{
C_3^{\rm vac.\ pol.}(e,\tau) \simeq
 {m_\mu^2 \over m_\tau^2} 
\left({4 \over 135} \ln{m_\mu\over m_e}-{1 \over 135}\right)
}
\nonumber \\
&&
\hspace*{-12mm}
+{m_\mu^4 \over m_\tau^4} \left( - {229213 \over 12348000}
          + {\pi^2 \over 630} 
          - {37 \over 11025} \ln{m_\tau\over m_\mu}
          - {1 \over 105} \ln{m_\tau\over m_\mu}
                          \ln{m_\tau m_\mu \over m_e^2}
          + {3 \over 4900} \ln{m_\mu\over m_e} \right) 
\nonumber \\
&& 
\hspace*{-12mm}
        +{m_\mu^6 \over m_\tau^6} \left( - {1102961 \over 75014100}
          + {4\pi^2 \over 2835} 
          - {398 \over 297675} \ln{m_\tau\over m_\mu}
          - {8 \over 945} \ln{m_\tau\over m_\mu}
                          \ln{m_\tau m_\mu \over m_e^2}
          - {524 \over 297675} \ln{m_\mu\over m_e} \right)
\nonumber \\
&&
 +  {2 \over 15} {m_e^2 \over m_\tau^2} 
 -{4\pi^2\over 45} {m_e^3\over  m_\tau^2 m_\mu}
\nonumber \\
&= & 0.0005276(2),
\label{eq:fin}
\ee 
in agreement with the numerical evaluation.  The error in the
result is due to the $\tau$--lepton mass uncertainty.  The
leading-logarithmic term of this expansion corresponds to simply
replacing $\alpha(q^2=0)$ by $\alpha(m_\mu^2)$ in the two-loop diagram
with a $\tau$ loop.  We have included the last term, with odd powers
of $m_e$ and $m_\mu$, even though it is not relevant numerically.  It
illustrates typical contributions of the eikonal expansion, the only
source of terms non-analytical in masses squared.  

We have checked eq.~(\ref{eq:fin}) by comparing it with an analytical
integration, using the kernel function given in \cite{Krause:1997rf}.
Terms quadratic in masses $m_{\tau,\mu,e}$ are in complete agreement.
The odd powers of $m_e$ have not been included in
\cite{Krause:1997rf}.

With formula (\ref{eq:fin}) the complete QED contribution to $a_\mu$
is now known analytically.  In this particular case this result does
not noticeably improve the accuracy of the QED prediction, since the
error in (\ref{eq:fin}) comes from the $\tau$ lepton mass
measurement.  However, the technique presented here might facilitate
other calculations.  We have seen that a combination of large
momentum, heavy mass, and eikonal expansions eliminates the need for
numerical calculations and enables us to construct arbitrarily
accurate expansions without knowledge of the exact result.

\section*{Acknowledgements}
We thank the CERN Theory Group for support and hospitality during our 
visit, where most of this project was completed.  
This work was supported in part by
the U.S.~Department of Energy
under grant number DE-AC02-98CH10886, 
Polish Government grants numbers
KBN 2P03B08414, 
KBN 2P03B14715, 
and
the Maria Sk\l{}odowska-Curie Joint Fund II PAA/DOE-97-316.


\begin{thebibliography}{10}


\bibitem{Bijnens:1996xf}
J. Bijnens, E. Pallante and J. Prades, Nucl. Phys. {\bf B474},  379  (1996).
\\
M. Hayakawa and T. Kinoshita, Phys. Rev. {\bf D57},  465  (1998).

\bibitem{Jeg95}
S. Eidelman and F. Jegerlehner, Z. Phys. {\bf C67},  585  (1995).
\\
M. Davier and A. H\"ocker, hep-ph/9805470.

\bibitem{CKM96}
A. Czarnecki, B. Krause and W. Marciano, Phys. Rev. Lett. {\bf 76},  3267
  (1996).

\bibitem{Czarnecki:1998nd}
A. Czarnecki and W. J. Marciano, hep-ph/9810512, talk given at 
  {\em 5th International Workshop on Tau Lepton Physics}, 
  Santander, Spain, September 1998. 

\bibitem{Schwinger48}
J. Schwinger, Phys. Rev. {\bf 73},  416  (1948).
\\
C.~M. Sommerfield, Phys. Rev. {\bf 107},  328  (1957);
 Ann. Phys. {\bf 5},  26  (1958).
\\
A. Petermann, Nucl. Phys. {\bf 3},  689  (1957);
 Helv. Phys. Acta {\bf 30},  407  (1957).

\bibitem{Laporta:1996mq}
S. Laporta and E. Remiddi, Phys. Lett. {\bf B379},  283  (1996).

\bibitem{Dehmelt87}
R. S. {van Dyck Jr.}, P. B. Schwinberg and H. G. Dehmelt,
Phys. Rev. Lett. {\bf 59}, 26 (1987).

\bibitem{Dubna}
T. Kinoshita, hep-ph/9808351, 
talk given at {\em International Workshop on Hadronic Atoms and Positronium
in the Standard Model}, Dubna, Russia, May 1998. 

\bibitem{Li:1993xf}
G. Li, R. Mendel and M.~A. Samuel, Phys. Rev. {\bf D47},  1723  (1993).

\bibitem{Elend66}
H.~H. Elend, Phys. Lett. {\bf 20},  682  (1966); erratum: ibid., {\bf
  21}, 720.

\bibitem{Laporta:1993pa}
S. Laporta and E. Remiddi, Phys. Lett. {\bf B301},  440  (1993).

\bibitem{Laporta:1993ju}
S. Laporta, Nuovo Cim. {\bf 106A},  675  (1993).

\bibitem{Krause:1997rf}
B. Krause, Phys. Lett. {\bf B390},  392  (1997).

\bibitem{Kinoshita:1993pq}
T. Kinoshita, Phys. Rev. {\bf D47},  5013  (1993).

\bibitem{Karshenboim:1993rt}
S.~G. Karshenboim, Phys. Atom. Nucl. {\bf 56},  857  (1993).


\bibitem{Chetyrkin91}
K.~G. Chetyrkin, preprint MPI-Ph/PTh 13/91.\\
V.~A. Smirnov, Mod. Phys. Lett. {\bf A10},  1485  (1995).\\
F.~V. Tkachev, Sov. J. Part. Nucl. {\bf 25},  649  (1994).

\bibitem{CzarSmir96}
A. Czarnecki and V.~A. Smirnov, Phys. Lett. {\bf B394},  211  (1997).

\bibitem{Smirnov96}
V.~A. Smirnov, Phys. Lett. {\bf B394},  205  (1997).

\end{thebibliography}

\end{fmffile}

\end{document}